\documentclass[fleqn,twoside]{article}
\usepackage[headings]{espcrc2}
\usepackage{graphicx}

\usepackage{amsmath}
\usepackage{amssymb}
\usepackage{amsthm}

\newcommand{\ZZ}{\mathbb{Z}_2}
\newcommand{\CC}{\mathbb{C}}

\newcommand{\Gr}{Gr\"obner }

\title{A Software Package to Construct Polynomial Sets
over $\ZZ$ for Determining the Output of Quantum Computations}

\author{Vladimir P. Gerdt\address[MCSD]{Laboratory of Information Technologies,
        Joint Institute for Nuclear Research, \\
        141980 Dubna, Russia}%
        \thanks{gerdt@jinr.ru},
        Vasily M. Severyanov\addressmark\thanks{severyan@jinr.ru}}

\runtitle{A Software Package to Construct Polynomial Sets over
$\ZZ$ for Quantum Computations} \runauthor{Vladimir Gerdt, Vasily
Severyanov}

\begin{document}

\begin{abstract}
A C\# package is presented that allows a user for an input quantum
circuit to generate a set of multivariate polynomials over the
finite field $\ZZ$ whose total number of solutions in $\ZZ$
determines the output of the quantum computation defined by the
circuit. The generated polynomial system can further be converted
to the canonical \Gr basis form which provides a universal
algorithmic tool for counting the number of common roots of the
polynomials.
\end{abstract}

\maketitle

\section{INTRODUCTION}

One important aspect of quantum computation is estimation of
computational power of quantum logical circuits. As it was
recently shown in~\cite{Dawson}, determining the output of a
quantum computation is equivalent to counting the number of
solutions of a certain set of polynomials defined over the finite
field $\ZZ$.

Using ideas published in~\cite{Dawson}, we have written a C\#
program enabling one to assemble an arbitrary quantum circuit
in a particular universal gate basis and to construct the
corresponding set of polynomial equations over
$\ZZ$. The number of solutions of the set defines the matrix
elements of the circuit and therefore its output value
for any input value.

The generated polynomial system can further be converted into the
canonical \Gr basis form by applying efficient involutive
algorithms described in~\cite{Gerdt}. A triangular \Gr basis for
the pure lexicographical order on the polynomial variables is
generally most appropriate for counting the number of common roots
of the polynomials.

Our program has a user-friendly graphical interface and a built-in
base of the elementary gates representing certain quantum gates and
wires. A user can easily assemble an input circuit from those
elements.

The structure of the paper is as follows. In
Section~\ref{sec:quantumcircuits} we outline shortly the circuit
model of quantum computation. Section~\ref{sec:someoverpaths}
presents the famous Feynman's sum-over-paths method applied to
quantum circuits. In Section~\ref{sec:decomposition} we describe
a circuit decomposition in terms of the elementary gates. In
Section~\ref{sec:assembling} we show how to assemble an arbitrary
circuit composed from the Hadamard and Toffoli gates that form a
universal basis. Section~\ref{sec:quantumpolynomials} demonstrates
a simple example of handling the polynomials associated with a
quantum circuit by constructing their \Gr basis.
We conclude in Section~\ref{sec:conclusions}.

\section{QUANTUM CIRCUITS}
\label{sec:quantumcircuits}

To quantize the classical bit, we go from the two-element set
$\{0,1\}$ to a two-level quantum system described by the
two-dimensional Hilbert space $\CC^2$.
In contrast to the classical case, the quantum bit (qubit)
can be found in a superposition $|\psi \rangle  = \alpha |0\rangle + \beta
|1\rangle$ of the  states $|0\rangle$ and $|1\rangle$ called a
computational basis, where $\alpha, \beta \in \CC$ are the
probability amplitudes of $|0\rangle$ and $|1\rangle$
respectively.

The simplest quantum computation is a unitary transformation on
the qubit state
$$
\left| \varphi  \right\rangle  = U\left| \psi
\right\rangle,\quad\quad\quad UU^\dag   = I.
$$
A measurement of the qubit in the computational basis $|0\rangle$
and $|1\rangle$ transforms its state to one of the basis states
with probabilities determined by the amplitudes
$$
\alpha |0\rangle  + \beta |1\rangle  \mapsto \left\{
{\begin{array}{*{20}c}
   {\left| 0 \right\rangle }\text{ with probability } |\alpha|^2 \\
   {\left| 1 \right\rangle }\text{ with probability } |\beta|^2  \\
\end{array}} \right.
$$

To compute a reversible Boolean vector-function $f:\ZZ^n\to
\ZZ^n$, one applies the appropriate unitary transformation $U_f$
to an input state $\left| \mathbf{a} \right\rangle$ composed of
some number of qubits
$$
\left| \mathbf{b} \right\rangle  = U_f \left| \mathbf{a}
\right\rangle, \quad\quad\quad \left| \mathbf{a} \right\rangle
,\left| \mathbf{b} \right\rangle \in \CC^{2\otimes n}
$$
The output state $\left| \mathbf{b} \right\rangle$ is not the
outcome of the computation until its measurement. After that
the output state can be used anywhere.

Some unitary transformations are called quantum gates. A
quantum gate acts only on a few qubits, on the rest it acts as the
identity. A quantum circuit can be assembled
by appropriately aligning quantum gates. The unitary transformation
defined by the circuit is the composition of the constituent unitary
transformations
\begin{equation}
U_f  = U_m U_{m - 1}  \cdots U_2 U_1
\label{eq:ufdecomposition}\end{equation}

A quantum gate basis is a set of universal quantum gates, i.e. any
unitary transformation can be presented as a composition of the
gates of the basis. As well as in the classical case, there are several
sets of universal quantum gates. For our work it is convenient to
choose the particular universal gate basis consisting of Hadamard
and Toffoli gates~\cite{Aharonov}.

The Hadamard gate is a one-qubit gate. It turns a computational
basis state into the equally weighted superposition
$$
\begin{array}{l}
 H:\left| 0 \right\rangle  \mapsto \frac{1}{{\sqrt 2 }}(\left| 0 \right\rangle  + \left| 1 \right\rangle ) \\
 H:\left| 1 \right\rangle  \mapsto \frac{1}{{\sqrt 2 }}(\left| 0 \right\rangle  - \left| 1 \right\rangle ) \\
 \end{array}
$$
The resulting superpositions for $\left| 0 \right\rangle$ and
$\left| 1 \right\rangle$ differ by a phase factor.

The Toffoli gate is a tree-qubit gate. Input bits $x$ and $y$ control
the behavior of bit $z$, and the Toffoli gate acts on
computational basis states as
$$
\left( {x,y,z} \right) \mapsto \left( {x,y,z \oplus xy} \right)
$$

An action of a quantum circuit can be described by a square unitary matrix
whose matrix element $\left\langle \mathbf{b} \right|U_f \left|
\mathbf{a} \right\rangle$ yields the probability amplitude for
transition from an initial quantum state $\left| \mathbf{a}
\right\rangle$ to the final quantum state $\left| \mathbf{b}
\right\rangle$. The matrix element is decomposed in accordance to
the gate decomposition of the circuit unitary
transformation~(\ref{eq:ufdecomposition}) and can be calculated as
sum over all the intermediate states $\mathbf{a}_i$, i = 1,2,
\ldots m - 1:
$$
\left\langle \mathbf{b} \right|U_f \left| \mathbf{a} \right\rangle
= \sum\limits_{\mathbf{a}_i } {\left\langle \mathbf{b} \right|U_m
\left| {\mathbf{a}_{m - 1} } \right\rangle } \cdots \left\langle
{\mathbf{a}_1 } \right|U_1 \left| \mathbf{a} \right\rangle
$$

\section{FEYNMAN'S SUM-OVER-PATHS}
\label{sec:someoverpaths}

\begin{figure}[b!]
\begin{center}
\includegraphics[width=7.4cm]{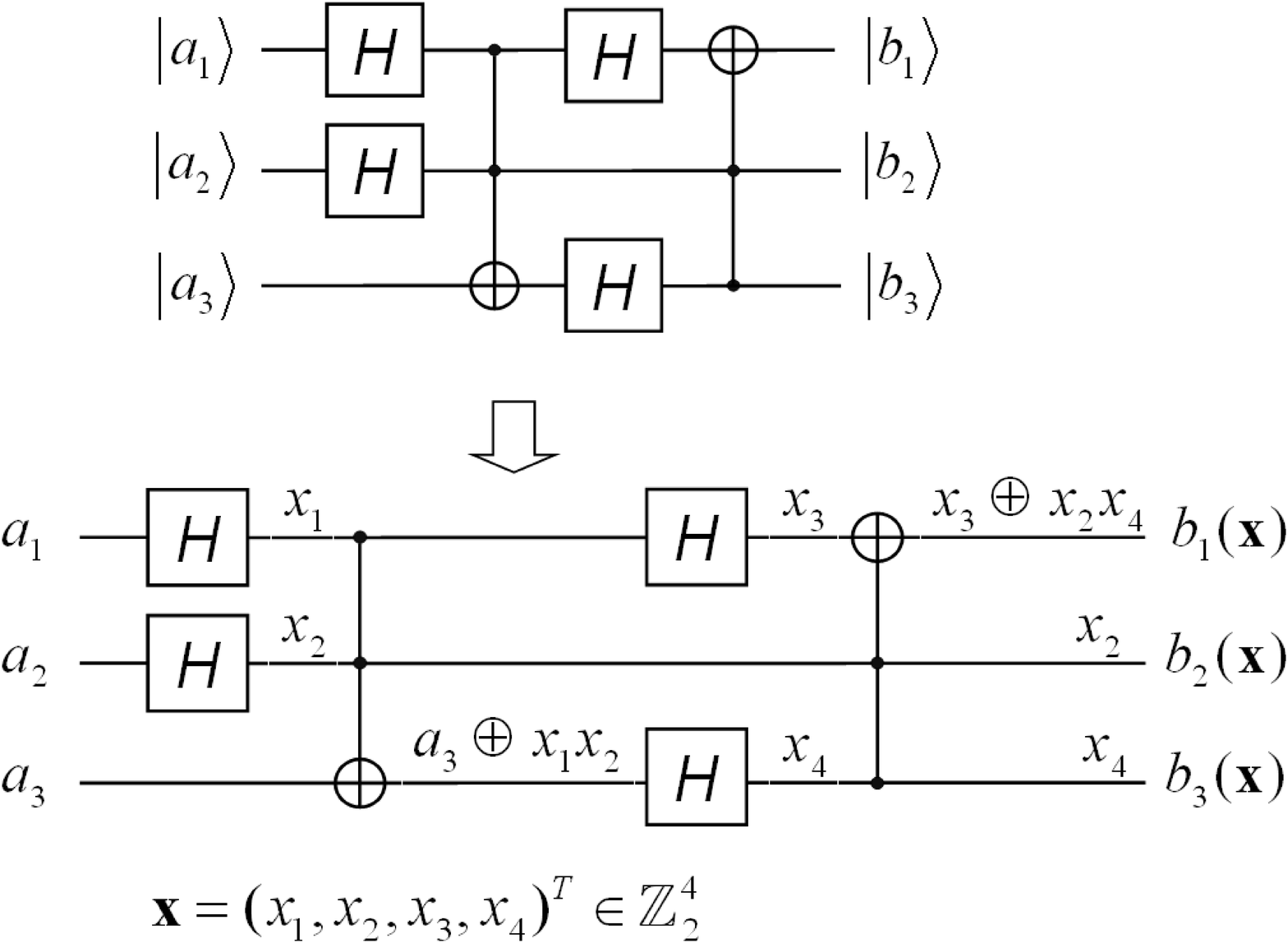}
\caption{From quantum to classical circuit}
\label{fig:quantum2classical}
\end{center}
\end{figure}

To apply the famous Feynman's sum-over-paths approach to calculate
the matrix element of a quantum circuit, we replace every quantum
gate of the circuit under consideration by its classical
counterpart. The trick here is to select the corresponding
classical gate for the quantum Hadamard gate because for any input
value, 0 or 1, it gives with equal probability either 0 or 1. We denote
the output of the classical Hadamard gate by the path variable $x$. Its
value determines one of the two possible paths of
computation. The classical Toffoli gate acts as
$$ \left( {a_1
,a_2 ,a_3 } \right) \mapsto \left( {a_1 ,a_2 ,a_3 \oplus a_1 a_2 }
\right),
$$
and the classical Hadamard gate as
$$
a_1  \mapsto x \qquad a_i ,x \in \ZZ
$$

Fig.~\ref{fig:quantum2classical} shows an example of quantum
circuit (taken from~\cite{Dawson}) and its classical
correspondence. The path variables $x_i$ comprise the (vector) path
$\mathbf{x} = (x_1 ,x_2 ,x_3 ,x_4 )^T  \in \ZZ^4$.

A classical path is a sequence of classical bit strings $a,a_1
,a_2 , \ldots ,a_m  = b$ resulting from application of the classical gates.
For each selection of values for the path variables
$x_i$ we have a sequence of classical bit strings which is called
an admissible classical path. Each admissible classical path has a
phase which is determined by the Hadamard gates applied. The phase is changed
only when the input and output of the Hadamard gate are simultaneously equal
to 1, and this gives the folmula
$$
\varphi (\mathbf{x}) = \sum\limits_{{\rm{Hadamard gates}}} {input
\bullet output}
$$
Toffoli gates do not change the phase.

For our example the phase of the path $\mathbf{x}$ is
$$
\varphi (\mathbf{x}) = a_1 x_1  \oplus a_2 x_2  \oplus x_1 x_3
\oplus x_4 (a_3  \oplus x_1 x_2 )
$$

The matrix element of a quantum circuit is given by sum over all the
allowed paths from the classical states $\mathbf{a}$ to $\mathbf{b}$
$$
\left\langle \mathbf{b} \right|U_f \left| \mathbf{a} \right\rangle
= \frac{1}{{\sqrt {2^h }
}}\sum\limits_{\mathbf{x}:\mathbf{b}\left( \mathbf{x} \right) =
\mathbf{b}} {\left( { - 1} \right)} ^{\varphi \left( \mathbf{x}
\right)}
$$
where $h$ is the number of Hadamard gates. The terms in the sum
have the same absolute value but vary in sign.

Let $N_0$ be the number of positive terms in the sum and $N_1$ the
number of negative terms
$$
N_0  = \left| {\left\{ {x|b(x) = b\quad \& \quad \varphi (x) = 0}
\right\}} \right|
$$
$$
N_1  = \left| {\left\{ {x|b(x) = b\quad \& \quad \varphi (x) = 1}
\right\}} \right|
$$
These equations count solutions to a system of $n+1$ polynomials in
$h$ variables over $\ZZ$. Then the matrix element may be
written as the difference
$$
\left\langle \mathbf{b} \right|U_f \left| \mathbf{a}
\right\rangle  = \frac{1}{{\sqrt {2^h } }}\left( {N_0  - N_1 }
\right)
$$

\section{CIRCUIT DECOMPOSITION}
\label{sec:decomposition}

\begin{figure}[h!]
\begin{center}
\includegraphics[width=7cm]{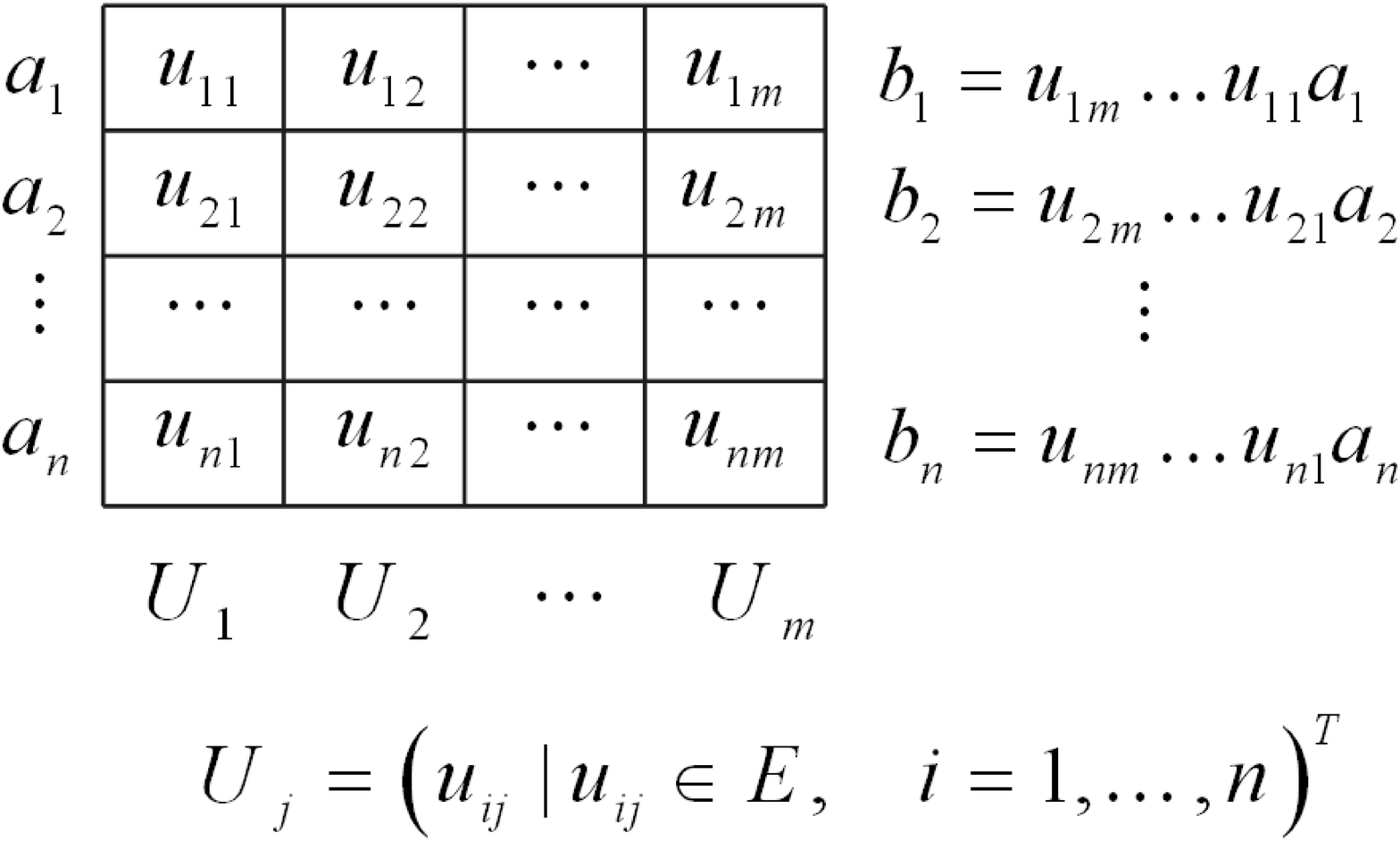}
\end{center}
\caption{Circuit decomposition into elementary gates}
\label{fig:decomposition}
\end{figure}

To provide a user with a tool for assembling
arbitrary quantum circuits composed from the Hadamard and Toffoli
gates we represent a circuit as a
rectangular table (Fig.~\ref{fig:decomposition}).

Each cell in the table
contains an elementary gate from following set
\begin{equation}
E = \{ I,\mathop {I,}\limits^ +  \mathop {I,}\limits^ \wedge
\mathop I\limits^ \vee  ,\mathop M\limits^ \wedge  ,\mathop
M\limits^ \vee  ,\mathop A\limits^ \wedge  ,\mathop A\limits^ \vee
,H\} \label{elementary gates}
\end{equation}
so that the output for each row is determined by the composition
of the elementary gates in the row. Thereby, each elementary unitary
transformation $U_j$ is represented as an n-tuple of elementary gates.

\begin{figure}[h!]
\begin{center}
\includegraphics[width=4cm]{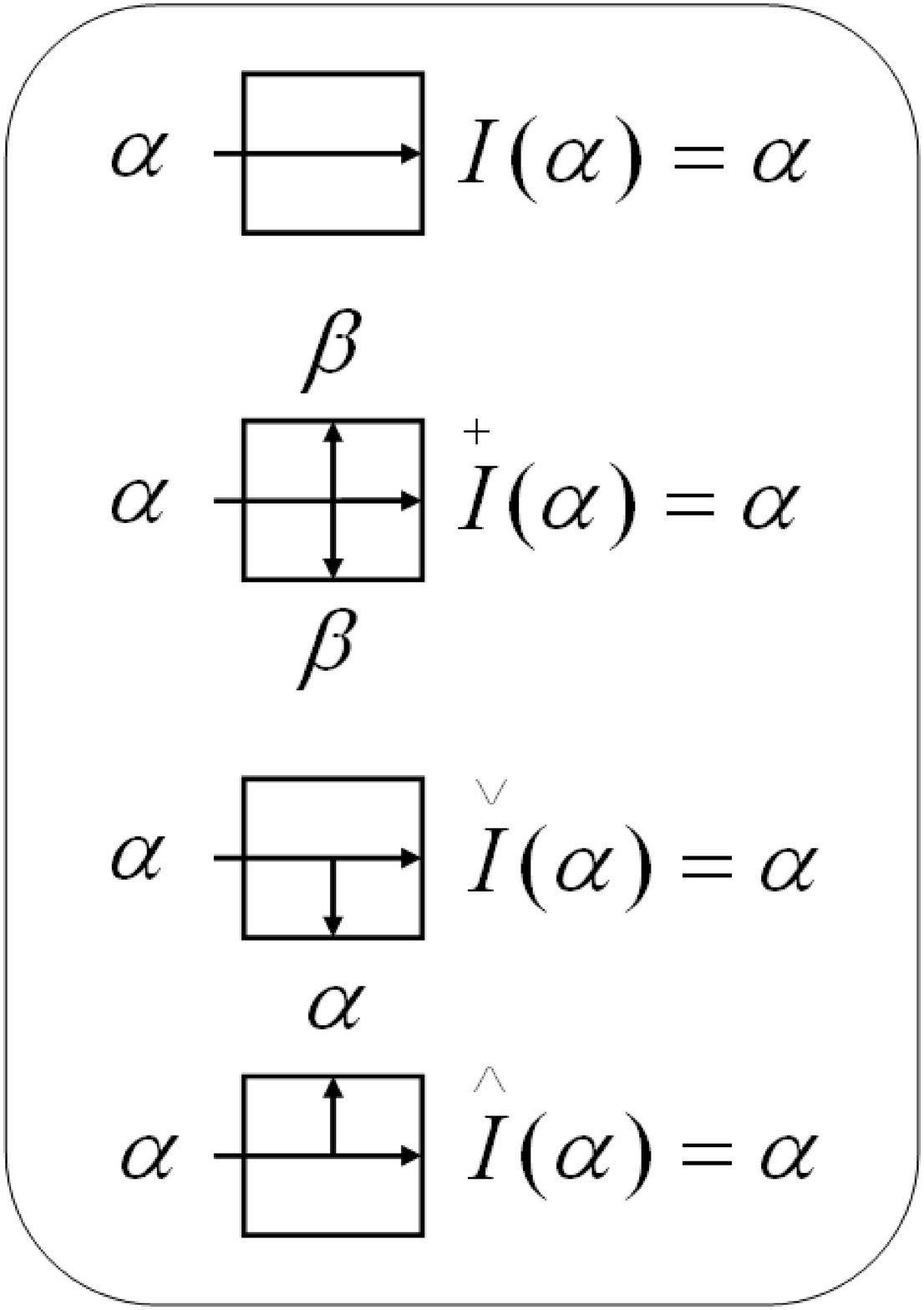}
\includegraphics[width=4cm]{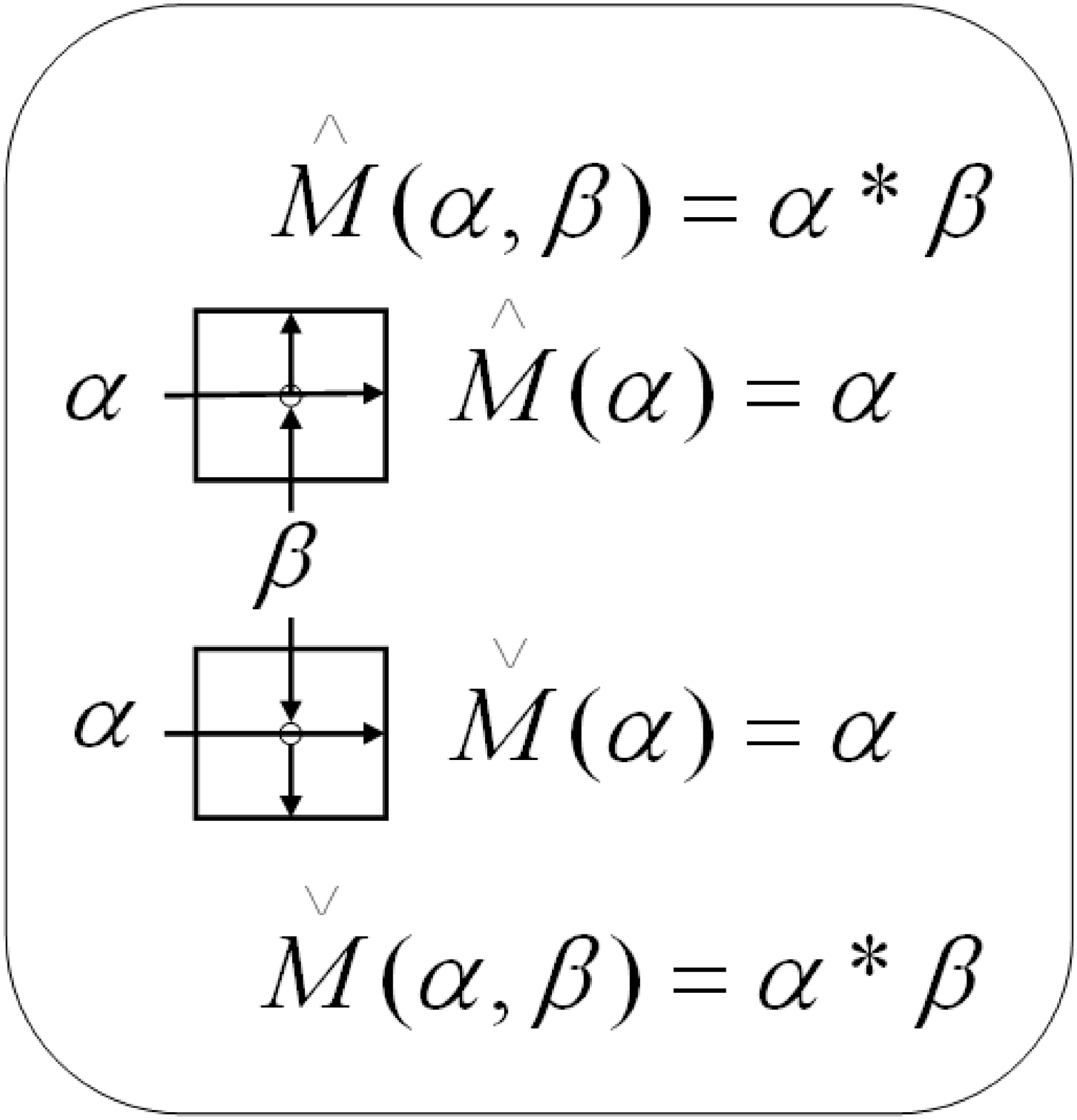}
\includegraphics[width=5cm]{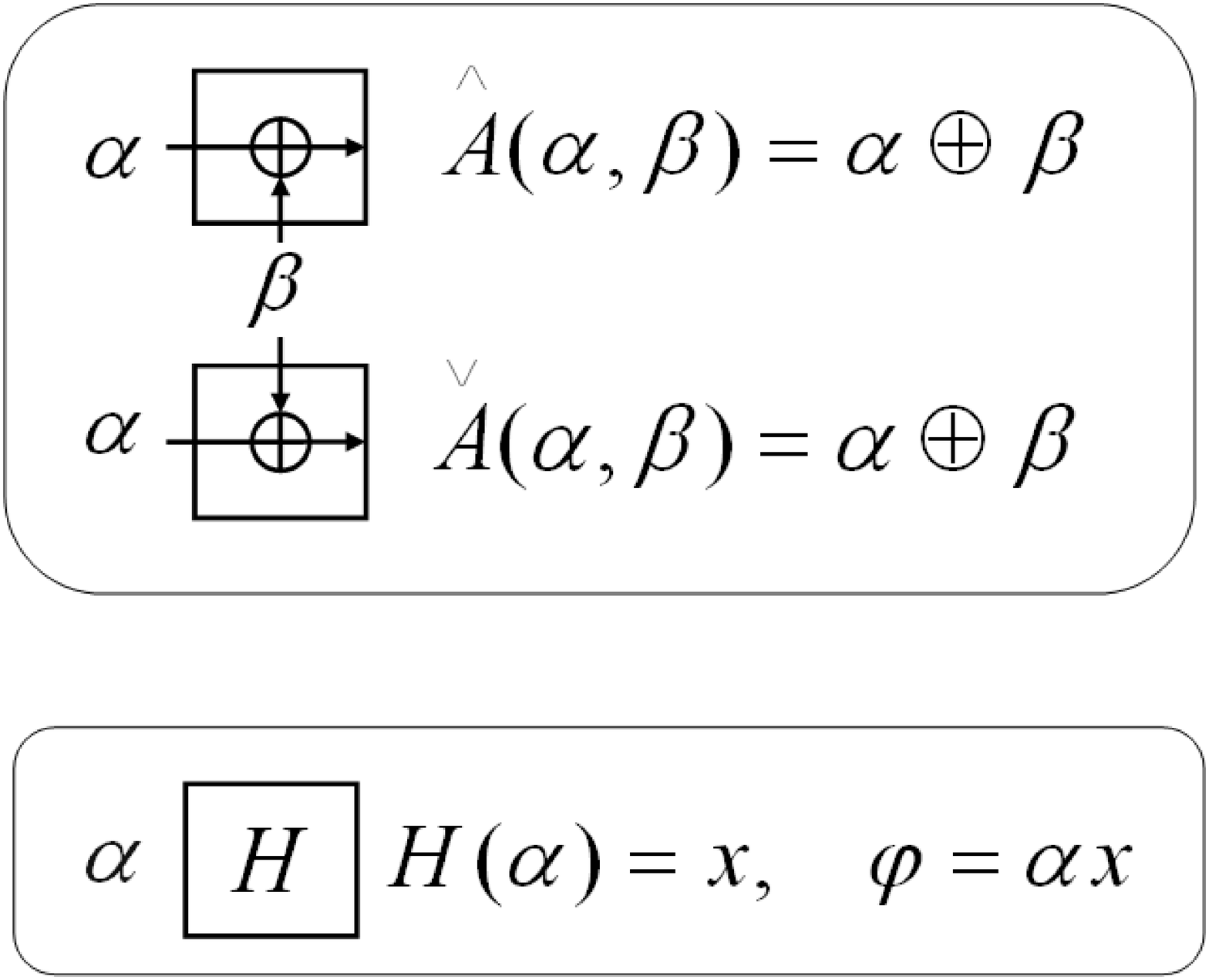}
\end{center} \caption{Action of elementary gates}
\label{fig:elelementarygates}
\end{figure}

Fig.~\ref{fig:elelementarygates} shows action of the elementary
gates from (\ref{elementary gates}): the identities, the multiplications, the additions
modulo 2, and the classical Hadamard gate. The identity
just
reproduces its input. The identity-cross
reproduces also its vertical input from the top elementary gate to the bottom one
and vice versa.
Every identity-down
and identity-up
have two outputs -- horizontal and vertical. The multiplication-up
and multiplication-down
perform multiplication of their horizontal and the corresponding
vertical inputs. In a similar manner act the addition-up
and addition-down.
Each Hadamard gate outputs an independent path variable irrespective of its input
and can give a nonzero contribution to the phase.

\section{ASSEMBLING CIRCUITS}
\label{sec:assembling}

\begin{figure}[h!]
\begin{center}
\includegraphics[width=7.4cm]{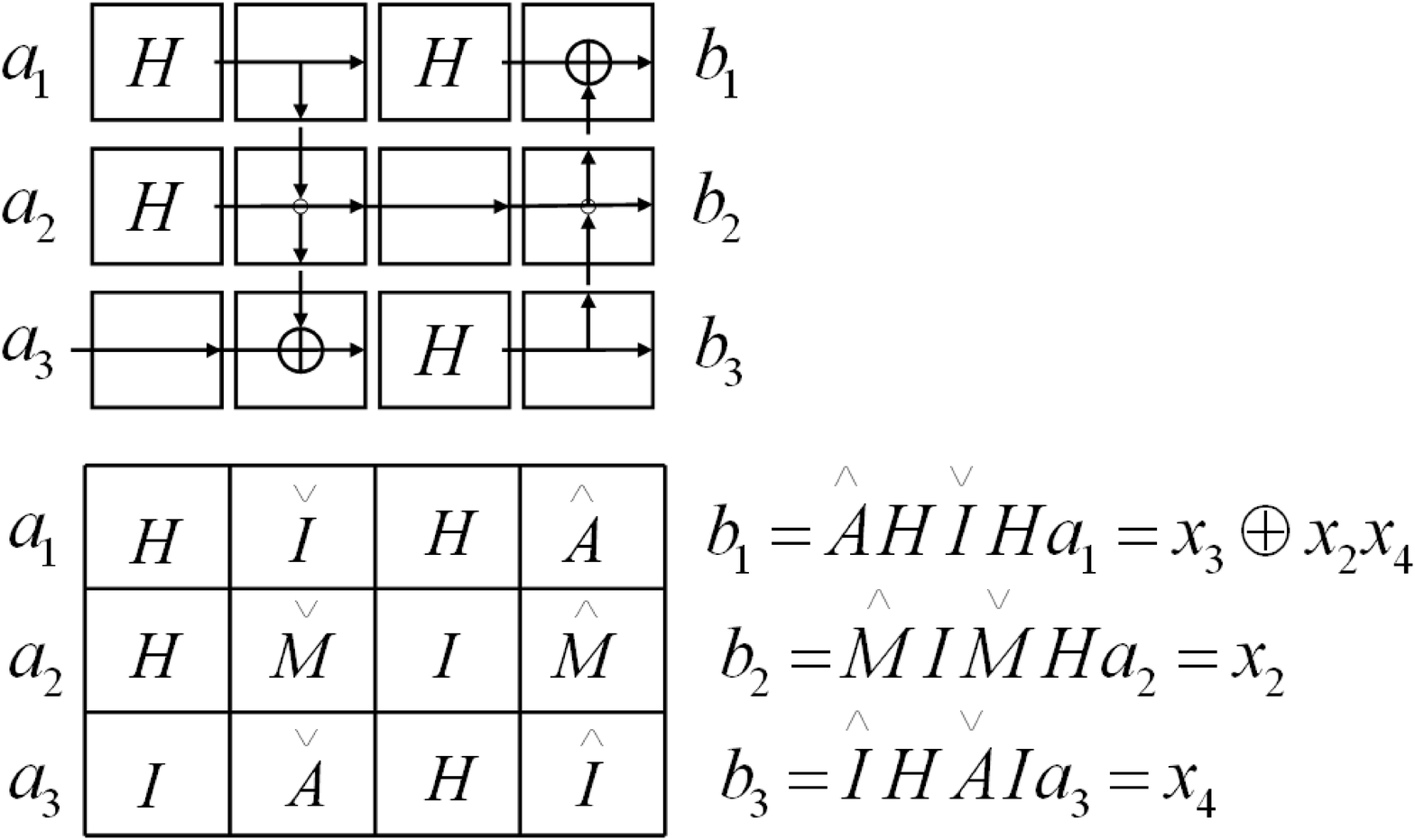}
\end{center}\caption{Assembling circuit} \label{fig:assembling}
\end{figure}

How can one assemble a circuit? First of all, we define an empty
table of the required size. In this case both output and phase are
not fixed. Then we place the required elementary gates in appropriate
cells. Now the output is the result of applying the elementary gates
to the input. The phase is also calculated. Then we proceed the same
way with the second column, with the third column, and so on up to
the last column. Fig.~\ref{fig:assembling} shows an example.

Apart from the ordinary menu, our program contains the toolbar for
selecting an elementary gate and the toolbar for main operations.
There are two windows: for assembling a circuit and for showing
its polynomials.

A circuit is represented in the program as two $2d$-arrays: one for
the elementary gates and another for their polynomials. The phase
polynomial is separately represented. The following piece of code
demonstrates construction of the circuit polynomials

\begin{verbatim}
  for each Column in Table of Gates
    for each Gate in Column {
      construct Gate Polynomial;
      if Gate id Hadamard
        reconstruct Phase Polynomial; }
\end{verbatim}
The method for constructing a gate polynomial is recursive because
of the need to go up or down for some gates.

Any circuit is saved as two files. One file is binary and contains the
circuit itself. Another file has a text format. It contains the circuit
polynomials in a symbolic form. The program allows to save
polynomials in several formats convenient for loading
into a computer algebra system (for example, in Maple or
Mathematica) for the further processing. It is also possible to load back
in memory a saved circuit.

Note, that the part of our code for name space {\bf Polynomial\_Modulo\_2}
written in C\#~\cite{Csharp} can also be used independently on our program.
This part contains classes for handling
polynomials over the finite field $\ZZ$. Class Polynomial is a list of
monomials, class Monomial is a list of letters, class Letter is an indexed
letter provided with a positive integer superscript (power degree).

\section{QUANTUM POLYNOMIALS}
\label{sec:quantumpolynomials}

A system generated by the program is a finite set $ F \subset \,R
$ of polynomials in the ring
$$
\begin{array}{l}
 R: = \ZZ [a_i ,b_j ][x_1 ,...,x_h ]\,\,\, \\
 \,a_i ,b_j  \in \,\,\ZZ ,\,\,\,i,j = 1,...n \\
 \end{array}
$$
in $h$ variables and $2n$ binary coefficients. One has to count
the number of roots $N_0$ and $N_1$  in  $\ZZ$ of the polynomial
sets
$$
F_0  = \{ f_{} ,...,f_k ,\varphi \,\} \,,\,\,F_1  = \{ f_{}
,...,f_k ,\varphi  + 1\,\} \,
$$
Then the circuit matrix is given by
$$
\left\langle {\mathbf{b} } \right|\,U\,\left| {\,\mathbf{a} }
\right\rangle = \frac{1}{{\sqrt {2^{\,h} } }}\,\,(\,N_0  - N\,_1
\,)
$$

To count the number of roots one can convert $F_0$ and $F_1$ into a
triangular form by computing the lexicographical Gr\"{o}bner basis
by means of the Buchberger algorithm or by involutive algorithm
decribed in \cite{Gerdt}.

For the example shown on Fig.~\ref{fig:quantum2classical} we
have the following polynomial system:
$$
\begin{array}{l}
 f_1  = x_2 x_4 \, + x_3  + b_1  \\
 f_2  = x_2  + b_2  \\
 f_3  = x_4  + b_3  \\
 \varphi  = \,x_1 x_2  + x_1 x_3  + a_1 x_1  + a_2 x_2  + a_3 x_4  \\
 \end{array}
$$
The lexicographical Gr\"{o}bner basis for the ordering $ x_1
\succ x_2 \succ x_3 \succ x_4 \,\,$ on the variables and representing both $F_0$ and
$F_1$ is as follows
$$
\begin{array}{l}
 g_1  = (a_1  + b_1 )x_1 \, + a_2 b_2  + a_3 b_3 \,\,( + 1) \\
 g_2  = x_2  + b_2  \\
 g_3  = \,\,x_3  + b_1  + b_2 b_3  \\
 g_3  = x_4  + b_3  \\
 \end{array}
$$
From this lexicographical Gr\"{o}bner basis
we immediately obtain the following conditions on the parameters:
$$
\begin{array}{l}
 a_1  + b_1  = 0\;\,\,\& \;\,\,\,a_2 b_2  + a_3 b_3  = 0 \\
 a_1  + b_1  = 0\;\,\,\& \;\,\,\,a_2 b_2  + a_3 b_3  = 1 \\
 \end{array}
$$
From these conditions we easily count 2 (0) roots of $F_0$ ($F_1$) and 0
(2) roots of $F_0$ ($F_1$). In all other cases there is 1 root of
$F_0$ and $F_1$.

Some matrix elements are
$$
\left\langle {000} \right|U\left| {001} \right\rangle  =  
\frac{1}{2}\,\,\,,\,\,\,\left\langle {000} \right|U\left| {111}
\right\rangle  = 0
$$

\section{CONCLUSION}
\label{sec:conclusions}

We presented the first version of a program tool for assembling
arbitrary quantum circuits and for constructing the corresponding
polynomial equation systems. Its number of solutions uniquely
determines the circuit matrix.

There is the algorithmic Gr\"{o}bner basis approach to converting
the system of quantum polynomials into a triangular form which is
useful for computing the number of solutions.

Thus, the above presented software together with \Gr bases provide a tool for
simulating quantum circuits.

\section{ACKNOWLEDGMENTS}

The research presented in this paper was partially supported by the grant 04-01-00784 from the Russian
Foundation for Basic Research.



\begin{thebibliography}{9}

\bibitem{Dawson} Christopher M. Dawson et al. {\em Quantum computing and polynomial
                 equations over the finite field $\ZZ$}. arXiv:quant-ph/0408129.

\bibitem{Gerdt} Gerdt V.P. {\em Involutive Algorithms for Computing Grobner Bases}.
Computational commutative and non-commutative algebraic
geometry, IOS Press, Amsterdam, 2005, pp.199-225.
arXiv:math.AC/0501111, 2005.

\bibitem{Aharonov} Aharonov D. {\em A Simple Proof that Toffoli and
Hadamard Gatesare Quantum Universal}. arXiv:quant-ph/0301040.

\bibitem{Csharp} {\em Microsoft Visual C\# .net Standard}, Version 2003.
\end{thebibliography}
\end{document}